\documentclass[12pt,hyper]{JHEP3}
\usepackage{graphicx,amssymb,amsbsy,amsmath,psfrag,empheq}

\newcommand{\nn}{\nonumber}
\newcommand{\no}{\noindent}
\newcommand{\ra}{\rightarrow} \newcommand{\om}{\omega}
\newcommand{\al}{\alpha} 

 \newcommand{\la}{\lambda}

\newcounter{multieqs}

\newcommand{\be}{\begin{equation}}
\newcommand{\ee}{\end{equation}}
\newcommand{\eq}[1]{(\ref{#1})}
\newcommand{\bit}{\begin{itemize}}  \newcommand{\eit}{\end{itemize}}

\def\nn{\nonumber}
\def\bea{\begin{eqnarray}}
\def\eea{\end{eqnarray}}

\def\one{\mbox{1 \kern-.59em {\rm l}}}

\def\d{\delta}    
\def\e{\epsilon}

\def\o{\omega}

\def\th{\theta}

  \def\cI{{\cal I}}

  \def\cR{{\cal R}}

\def\Tt{\tilde{T}}

\def\d{\delta}

\def\uno{\mbox{1 \kern-.59em {\rm l}}}

\def\one{1\!\!1\,\,}

\def\bcomment#1{}

\def\rC { \textrm{C} }
\def\rD { \textrm{D} }
\def\rT { \textrm{T} }
\def\rP { \textrm{P} }
\def\rNP { \textrm{NP} }
\def\rR {\textrm{\cR}}

\def\IR{\relax{\rm I\kern-.18em R}}

\title{Nonequilibrium Dynamics in Noncommutative Spacetime}

\author{Chong-Sun Chu$^a$ and Chiu Man Ho$^b$ \\
$^a$Centre for Particle Theory
and Department of Mathematics,
Durham University, Durham, DH1 3LE, UK \\
$^b$Department of
  Physics, University of California, Berkeley, CA 94720, USA\\
Theoretical Physics Group, Lawrence Berkeley National
  Laboratory, Berkeley, CA 94720, USA\\
Department of
  Physics and Astronomy, Vanderbilt University, Nashville, Tennessee
  37235, USA\\
\\{\tt chong-sun.chu@durham.ac.uk,}\,
{\tt chiuman.ho@vanderbilt.edu} }

\abstract{
We study the effects of spacetime noncommutativity on the
nonequilibrium dynamics of particles in a thermal bath.
We show that the noncommutative thermal bath does not
suffer from any further 
IR/UV mixing problem in the sense that all the finite-temperature non-planar
quantities are free from infrared singularities.
We also point out that
the combined effect of finite temperature and
noncommutative geometry has a distinct effect on
the nonequilibrium dynamics of particles propagating in a thermal bath:
depending on the momentum of the mode of concern,
noncommutative geometry may switch on or switch off
their decay and thermalization.
This
could have
significant impacts on the nonequilibrium phenomena in
the early universe at
which spacetime
noncommutativity may be present.
Our results suggest a re-examination of some of the
important processes in the early universe such as
reheating after inflation, baryogenesis
and the freeze-out of superheavy dark matter candidates.}

\preprint{DCPT-09/83}

\begin{document}

\section{Introduction}

On the noncommutative (NC) spacetime,  the spacetime coordinates do
not commute with each other anymore but obey the commutation relation:
\be
[x^{\mu},\,x^{\nu}]= i\theta^{\mu\nu}
\ee
where $\theta^{\mu\nu}$
is a constant antisymmetric matrix \cite{Alain}.  Noncommutative quantum
field theory (NCQFT) can then be derived from its commutative
counterpart with the usual product of fields replaced by the Moyal
star product:
\be
(\phi\star\chi)(x) \equiv e^{\frac{i}{2}\,
  \theta^{\mu\nu}
  \partial^x_{\mu}\partial^y_{\nu}}\,\phi(x)\,\chi(y)|_{y=x}.
\ee
As
NCQFT arises naturally from string theory
\cite{ChuHo,Schomerus,SeibergWitten}, it has received lots of
attentions and has been an active research topic during the last
decade.
NCQFT has many unique properties such as Lorentz
violation \cite{Kostelecky}, nonlocality and modified causality
\cite{AGVM,CFI}.
Another intriguing phenomenon associated with
NCQFT is the existence of the  
infrared (IR)/ultraviolet (UV)  singularities 
\cite{SeibergPerturbative,Raamsdonk,Susskind}.  This is a
phenomenon which gives rise to various pathologies in the field
theory.  Despite the 
loss of Lorentz invariance and locality, it has
been argued that both CPT and spin-statistics theorems still hold
\cite{Sheikh,Wess,Chaichian}. However, it has been pointed out that
the  space-time noncommutative theory (i.e. $\theta^{0i}\neq 0$ for $i
=1,2,3$) may violate unitarity \cite{Gomis} if the theory also suffers
from 
IR/UV mixing \cite{Zwicky,CLZ}.  Therefore, to avoid getting into
trouble with unitarity, we will confine ourselves to the case with
$\theta^{0i}= 0$ in our study.

On the other hand, nonequilibrium phenomena play a crucial role in
many important processes in the early universe. These include
reheating after inflation, baryogenesis, freeze-out of dark matter
candidates, electroweak and QCD phase transitions
\cite{kolbturner,bernstein,dodelson}. A common treatment of
nonequilibrium evolution is to implement the closed-time-path (CTP)
formalism \cite{ctp1,ctp2,ctp3,ctp4} which is a path-integral approach
to a time evolved density matrix.  The thermal bath degrees of freedom
are integrated out to obtain the nonequilibrium effective action which
forms the generating functional for all the correlation
functions. This approach also leads to quantum Boltzmann equations
which can be solved to give the time evolution of the distribution
functions. The accomplishment of thermal equilibrium is determined by
the asymptotic time behaviour of the equal-time two-point correlation
function and the distribution function.

A natural question to be asked would be: what if we consider finite
temperature and spacetime noncommutativity at the same time?
Since noncommutative geometry naturally introduces a new 
energy scale
$E_{\rm NC} \sim \th^{-1/2}$ in addition to the temperature scale $T$,
noncommutativity could have an  interesting impact on the time
evolution of a nonequilibrium system.
In particular, if spacetime noncommutativity is really present in the early
universe, it would be important to understand how does it affect the
relevant physics. 
Some behaviours of NCQFT at finite
temperature have been investigated in
\cite{Arcioni,BarbonGomis,Fischler,Paban,Landsteiner1,Landsteiner2,TE1,TE2}.
However, none of these works have
considered the decay and
thermalization of particles propagating in a noncommutative thermal
bath.  In order to fill the gap,
in this paper we raise and investigate the following question: 
how does spacetime noncommutativity affect the nonequilibrium dynamics?

In order to address this question,  we
consider a simple model
with scalar $\Phi$ particles propagating in a thermal bath constituted
by two other different scalars $\chi_1$ and $\chi_2$. The question of
how do the $\Phi$ particles come from at the first place is irrelevant
to our discussion.
Immediately after
the $\Phi$ particles are created, they are
not in equilibrium
with the
thermal bath.  They can either decay into $\chi_1+\chi_2$ or
thermalize with them, or both, depending on their kinematical
properties. A \emph{complete} understanding of the nonequilibrium
dynamics of $\Phi$ would be to study the time evolution of their
correlation and distribution functions. However, as a first step, we
will only confine ourselves to their decay and thermalization
processes in this study. The decay rate $\Gamma_\textrm{D}$ and
thermalization rate $\Gamma_\textrm{T}$ of the $\Phi$ particles
are characterized by the imaginary part of their self-energy as well as
their in-medium dispersion 
relation \cite{CMHo}. Notice that in the
expanding early universe, even if $\Gamma_\textrm{T} \neq 0$ and
thereby a thermalization process is kinematically favored, it still
does not guarantee an actual thermalization. To maintain thermal
equilibrium with the thermal bath in the early universe, we require
$\Gamma_\textrm{T}> H$ where $H$ is the Hubble expansion rate. If
$\Gamma_\textrm{T} = 0$  or $\Gamma_\textrm{T} < H$, then the $\Phi$
particles will continue to be nonequilibrium; and whether they can
have out-of-equilibrium decays depends on $\Gamma_\textrm{D}\neq 0$ or
$\Gamma_\textrm{D} = 0$.

This article is organized as follows. In Section 2, we will outline
our model and sketch how to compute the different contributions to the
imaginary part of the self-energy, which will be analyzed to reveal its
properties.
We will also compute the real part of the self-energy and hence
obtain the noncommutative
in-medium dispersion relation.
The  IR/UV mixing issue in our model will be analyzed as well.
In Section 3, we will study the
impacts of spacetime noncommutativity on the decay and thermalization
processes of the 
$\Phi$
particles propagating in the thermal bath.
We will find that as a combined result of
finite temperature and noncommutative geometry,
the stability and
thermalizability properties of the 
$\Phi$
particles propagating in the thermal bath
are altered in a momentum dependent manner.
Finally, in Section 4, we will give some preliminary
discussions on the possible applications of our results
to the early universe.

\section{The Model}

We consider a theory of three interacting real scalar fields in
noncommutative spacetime with the following Lagrangian density
\bea \label{Lag} \mathcal{L}= \frac{1}{2}
\partial_{\mu}\Phi\partial^{\mu}\Phi - \frac{1}{2} M_B^2 \Phi^2 +
\sum_{i=1}^2 \Bigg[\frac{1}{2}
  \partial_{\mu}\chi_i\partial^{\mu}\chi_i - \frac{1}{2} M^2_i
  \chi^2_i \Bigg]-\frac{g}{2}\,\chi_1\star \Phi \star \chi_2
-\frac{g}{2}\,\chi_2\star \Phi \star \chi_1\, , \;\;\;\;\;\;\;
\eea
where $M_B$ is the bare mass of $\Phi$.  We will assume that the
mutual interaction between the fields $\chi_1\,,\,\chi_2$  ensures
them to be in thermal equilibrium at a temperature $T=1/\beta$. The
commutative counterpart of this model has been previously studied in
\cite{CMHo,weldon}
with an analysis of the different processes in the
thermal medium. Here we will follow the similar treatment and
conventions as \cite{CMHo}.

The relevant quantity is the self-energy of the field $\Phi$ which we
will obtain to one loop order $\mathcal{O}(g^2)$ in the Matsubara
representation. The one-loop self-energy is given by
\be\label{sigmafi}
\Sigma(\nu_{n}, \vec{k})= -g^{2}\int
\frac{d^{3}\vec{p}}{(2\pi)^{3}}\frac{1}{\beta}\sum_{\omega_{m}}G_{\chi_{1}}
(\omega_{m},  \vec{p})\,G_{\chi_{2}}(\omega_{m}+\nu_{n},
\vec{p}+\vec{k})\, \left(\,\frac12 +\frac12 \,e^{ip\times k}\,\right),
\end{equation}
where $ \omega_m= 2\pi m/\beta,\,\nu_n=2\pi n/\beta$ are the
bosonic  Matsubara frequencies and
$p\times k\equiv p_{i}\,\theta^{ij}\,k_{j}$.  The factors $\frac12$
and $\frac12 \,e^{ip\times k}$ represent the planar and non-planar
contributions respectively.
Obviously the noncommutative phase factor is nontrivial only if $k^i$
is  nonvanishing in the direction where $\theta^{ij}$ is non-zero .
The Matsubara propagators $G_{\chi_{1}}$ and $G_{\chi_{2}}$ are
written in the following dispersive form
\begin{eqnarray}
G_{\chi_{1}}(\omega_{m}, \vec{p})&=& \int dp_{0}\frac{\rho_{1}(p_{0},
  \vec{p})}{p_{0}-i \omega_{m}}, \\ G_{\chi_{2}}(\omega_{m}+\nu_{n},
\vec{p}+\vec{k})&=& \int dq_{0}\frac{\rho_{2}(q_{0},
  \vec{p}+\vec{k})}{q_{0}-i \omega_{m}-i\nu_{n}}\,,
\eea
where the spectral densities for $\chi_1$ and $\chi_2$ are
\bea
\rho_{1}(p_{0},\vec{p})&=& \frac{1}{2\,\omega_1}
     [\delta(p_{0}-\omega_1)-\delta
       (p_{0}+\omega_1)]\, , \quad
     \omega_1= \sqrt{\vec{p}^{2}+M_{1}^{2}},
     \\ \rho_{2}(q_{0},
     \vec{p}+\vec{k})&=&\frac{1}{2\,\omega_2}[\delta
       (q_{0}-\omega_2)-\delta
       (q_{0}+\omega_2)]\, , \quad
     \omega_2 =
     \sqrt{(\vec{p}+\vec{k})^{2}+M_{2}^{2}}. \,\;\;
\eea
This representation allows us to carry out the sum over Matsubara
frequencies $\omega_{m}$ in a rather straightforward manner
\cite{kapusta,lebellac}. The resulting self-energy
can be further
written in the dispersive form
\be
\label{sigdis}
\Sigma(\nu_n,\vec{k}) = -\frac{1}{\pi} \int_{-\infty}^{\infty} d\omega
\frac{\mathrm{Im} \Sigma_\textrm{R}(\omega,k)}{\omega-i \nu_n},
\ee
where $\text{Im} \Sigma_\textrm{R}(\omega,k)$ is the imaginary part of
the retarded self-energy which
is defined by the analytic continuation
\be
\Sigma_\textrm{R}(k_0,k) = \Sigma(\nu_n=-i k_0-\epsilon, k).
\ee

\subsection{Imaginary part of the self-energy}

The retarded self-energy  $\Sigma_\rR(\om)$ has cuts along the real axis.
The discontinuity across these cuts (defined by
$\Sigma_\rR(\om + i \e) -  \Sigma_\rR(\om - i \e)$)
gives the imaginary part
of  $\Sigma_\rR(\om)$:
\be
{\rm Disc}\, \Sigma_\rR(\om) = - 2i\, {\rm Im}\Sigma_\rR(\om).
\ee
It is then easy to obtain
\begin{empheq}[left={
{\rm Im}\Sigma_\rR(\om, \vec{k}) = g^2  \int \dfrac{d^3 \vec{p}}{(2\pi)^3}
\dfrac{2 \pi}{2 \om_1 \, 2 \om_2} \empheqlbrace},
right = {\empheqrbrace \big(
\frac{1}{2} + \frac{1}{2} e^{i p \times k}\big),} ]{align}
&
\big[\d(\om-\om_1- \om_2) - \d(\om + \om_1+ \om_2)\big] \nn\\
&  \qquad \qquad \qquad \qquad \qquad \cdot(1+ n_1+ n_2)\nn\\
& ~+ \d(\om+\om_1 -\om_2) (n_1 - n_2)\nn\\
 & ~+ \d(\om+\om_2 -\om_1) (n_2 -  n_1) \nn
\end{empheq}
where
\bea
n_i = n(\om_i)~~,~~n(\om) = \frac{1}{e^{\om/T} -1}.
\eea
Here the factor $\frac{1}{2}$ gives the planar contribution
$\text{Im}\Sigma^{\textrm{P}}_\textrm{R}$, while the
factor $\frac{1}{2} e^{ p \times k}$ gives  the nonplanar contribution
$\text{Im}\Sigma^{\textrm{NP}}_\textrm{R}$. We will use
the superscripts ``P" and ``NP" denote the planar and
non-planar contributions respectively.

It is convenient to write $\text{Im}\Sigma^{\textrm{P,NP}}_\textrm{R}$
as a sum of several contributions of different physical origin, namely
\begin{eqnarray}\label{imsigsplit}
\text{Im}\Sigma^{\textrm{P}}_\textrm{R}(\omega,k,T)&=&
\sigma_{0}^{\textrm{P}}(\omega,k)+\sigma_{a}^{\textrm{P}}(\omega,k,T)
+\sigma_{b}^{\textrm{P}}(\omega,k,T)\;,\\
\text{Im}\Sigma^{\textrm{NP}}_\textrm{R}(\omega,k,T)&=&
\sigma_{0}^{\textrm{NP}}(\omega,k)+\sigma_{a}^{\textrm{NP}}(\omega,k,T)
+\sigma_{b}^{\textrm{NP}}(\omega,k,T)\;.
\end{eqnarray}
Here the quantities
$\sigma_0^{\textrm{P,\,NP}}(\omega, k)$ are the  zero-temperature
contributions,  while $\sigma^{\textrm{P,\,NP}}_a(\omega, k,T)$,
$\sigma^{\textrm{P,\,NP}}_b(\omega, k,T)$ are the finite-temperature
contributions.  At one-loop,
the processes that contribute to
$\sigma^{\textrm{P,\,NP}}_0(\omega, k)$,
$\sigma^{\textrm{P,\,NP}}_a(\omega, k,T)$ are $\Phi \leftrightarrow
\chi_1 + \chi_2$, while the processes that contribute to
$\sigma^{\textrm{P,\,NP}}_b(\omega, k,T)$ are $\chi_{1,2}
\leftrightarrow \Phi+ \chi_{2,1}$. See Figure 1.

\psfrag{chi1}{$\chi_1, \vec{p}$}
\psfrag{chi2}{$\chi_2, \vec{p}+ \vec{k}$}
\psfrag{phi}{$\Phi, \vec{k}, \omega$}
\begin{figure}[bpt]
\label{fig1}
\includegraphics[scale=.65]{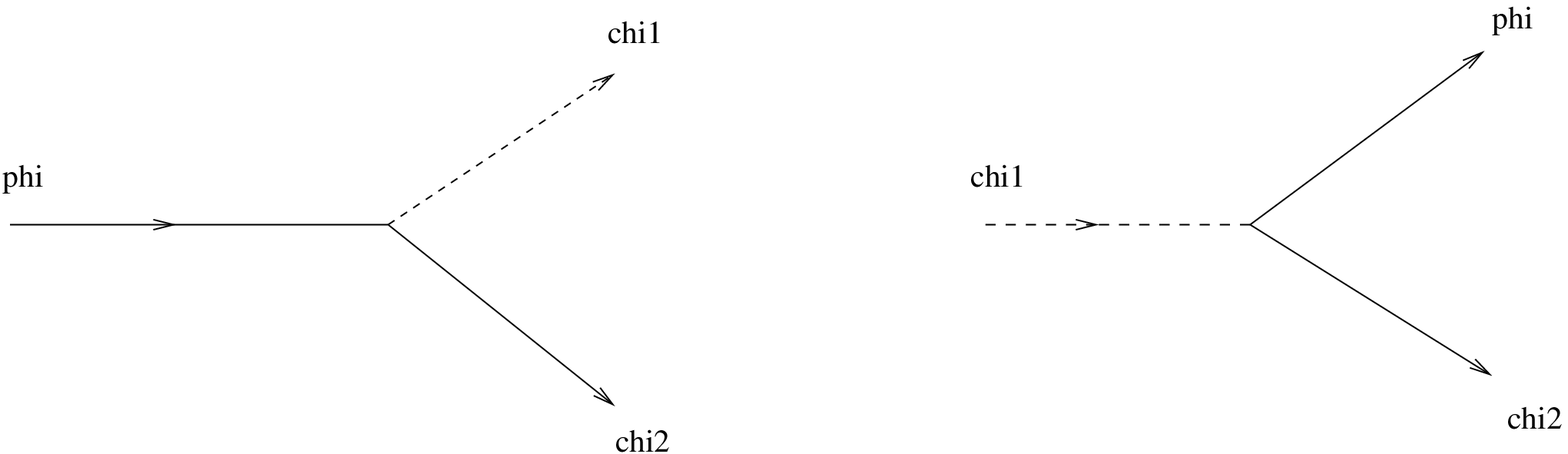}\\
{Fig 1: One-loop processes contributing to
$  \sigma^{\textrm{P,\,NP}}_0(\omega, k),
  \sigma^{\textrm{P,\,NP}}_a(\omega, k,T)$  and
 $\sigma^{\textrm{P,\,NP}}_b(\omega, k,T)$.  The inverse processes
  are not shown.}
 \end{figure}

In \cite{CMHo}, the \textit{commutative} version of our current model
has been studied and the corresponding contributions to the imaginary
part of the self-energy are computed as $\sigma_0, \sigma_a$ and
$\sigma_b$ respectively. In fact,
$\text{Im}\Sigma_\textrm{R}^{\textrm{P}}$ is precisely $\frac12$ times
the corresponding results in \cite{CMHo}, which implies that
$\sigma_0^{\textrm{P}}(\om,k)=\frac12\,\sigma_{0}$,
$\sigma_{a}^{\textrm{P}}(\omega,k,T)  = \frac12 \,\sigma_a$,
$\sigma_{b}^{\textrm{P}}(\omega,k,T)  =\frac12\,\sigma_b$.  We have:
\bea
\sigma_0^{\textrm{P}}(\om,k)&=& \frac{g^{2}}{32 \pi k}\,
\text{sign}(\omega)\, \Theta[\,Q^2-(M_{1}+M_{2})^{2}\,]\,
(\,B-A\,)\,, \label{Si1a}\\
\sigma_{a}^{\textrm{P}}(\omega,k,T)  &=&
\frac{g^{2}}{32 \pi \beta k }\, \text{sign}(\omega)\,
\Theta[\,Q^2-(M_{1}+M_{2})^{2}\,]\,\bigg[ \ln \left(\, \frac{
    1-e^{-\beta B }}{ 1-e^{-\beta A}}\,\right)+ (M_1 \leftrightarrow
  M_2) \bigg],\;\;\;\;\;\;\;\;\;\; \label{Si2a}\\
\sigma_{b}^{\textrm{P}}(\omega,k,T)
&=& \frac{g^{2}}{32 \pi \beta k }\, \text{sign}(\omega)\,
\Theta[\,(M_{1}-M_{2})^{2}-Q^2\,]\,\bigg[ \ln \left( \,\frac{
    1-e^{-\beta A} }{ 1-e^{-\beta B}}\,\right)+ (M_1 \leftrightarrow
  M_2) \bigg],\;\;\;\;\;\;\;\; \label{Si3a}
\eea
where $A = |\omega_p^{-}|\,,\,B = |\omega_p^{+}|$,
\be
Q^2 =\om^2-k^2
\ee
and
\begin{equation}
\omega_{p}^{\pm}=\frac{|\omega|\,[\,Q^2+(M_1^2-M_2^2)\,]\pm
  k\sqrt{[\,Q^2+(M_1^2-M_2^2)\,]^{2}-4\, Q^2\, M_{1}^{2}}}{2\,Q^2}\,.
\label{omegappm}
\end{equation}
In $\sigma_a^{\textrm{P}}$ and $\sigma_b^{\textrm{P}}$, $\beta k \ll
1$ corresponds to the high temperature limit while $\beta k \gg 1$
corresponds to the low temperature limit. Both $\sigma_a^{\textrm{P}}$
and $\sigma_b^{\textrm{P}}$ increases with decreasing $\beta k$. For
instance, $\sigma^{\textrm{P}}_{a,b} \ra 0$ when $\beta k \ra
\infty$. On the other hand, both $\sigma_a^{\textrm{P}}$ and
$\sigma_b^{\textrm{P}}$ approach to a finite value when $\beta k \ra
0$.

As for the non-planar parts,  we have
to include
the
factor $e^{ip\times k}$ in the integral. To evaluate the integral, we
note that since the vector $\theta^{ij} k_j$ is perpendicular to
$k^i$, it is convenient to adopt a spherical coordinate system with
the polar axis pointing in the $k^i$ direction and the azimuthal angle
$\phi$ measured from an axis defined by $\theta^{ij} k_j$. Denoting
the polar angle by $\vartheta$, we obtain
\be
p\times k = p \,\theta \, k \,\sin \vartheta \cos \phi,
\ee
where $\theta$ is the magnitude
of the vector $\theta^{ij} \hat{k}_j$ and $\hat{k}^i$ is the unit
vector of $k^i$. For example, if $k^i=(0,0,k)$, then $\theta \equiv
\sqrt{(\theta^{31})^2+ (\theta^{32})^2}$.  The calculations then
proceed in a similar fashion as has been performed in the appendix of
\cite{CMHo}.  After lengthy but straightforward calculations, we
obtain the non-planar contributions to the imaginary part of the
self-energy
\begin{eqnarray}
\sigma_{0}^{\textrm{NP}}&=&\frac{g^{2}}{16 \pi}\,
\text{sign}(\omega)\, \Theta[\,Q^2-(M_{1}+M_{2})^{2}\,]\,\, \frac{1}{k
  Q\,\theta}\,\sin\left(\Omega \,k Q\,\theta\right),
\label{Si1b}
\\
\sigma_{a}^{\textrm{NP}}&=&\frac{g^{2}}{16 \pi}\,
\text{sign}(\omega) \,\Theta[\,Q^2-(M_{1}+M_{2})^{2}\,]\,\,
\sum_{n=1}^{\infty}\,\frac{e^{-\frac{B+A}{2k}\,\beta k\,n}}{\beta
  k\,\sqrt{(\frac{Q\,\theta}{\beta})^2-n^2}} \,\sin\bigg(\Omega \, \beta
k\,\sqrt{\left(\frac{Q\,\theta}{\beta}\right)^2-n^2}\bigg)
\nonumber
\\
&& + \left(\,M_1 \leftrightarrow M_2 \,\right),
\label{Si2b}
\\
\sigma_{b}^{\textrm{NP}} &=&-\frac{g^{2}}{16\pi
}\,\text{sign}(\omega) \left[\Theta(-Q^2)\,\,
  \sum_{n=1}^{\infty}\frac{e^{-\frac{B+A}{2k}\,\beta k
      \,\sqrt{(\frac{Q\,\theta}{\beta})^2+n^2}}} {\beta
    k\,\sqrt{(\frac{Q\,\theta}{\beta})^2+n^2}} \,\sinh\left(\Omega \,\beta
  k\;n \right)\right.
\nonumber
\\
&& ~~+\left. \,\Theta(Q^2)
  \,\Theta[\,(M_{1}-M_{2})^{2}-Q^2\,] \,\, \sum_{n=1}^{\infty}\,
  \frac{e^{-\frac{B+A}{2k}\,\beta k \,n}} {\beta
    k\,\sqrt{(\frac{Q\,\theta}{\beta})^2-n^2}} \,\sin\bigg(\Omega \,\beta
  k\,\sqrt{\left(\frac{Q\,\theta}{\beta}\right)^2-n^2}\bigg)\right]\nonumber
\\ &&~~ + \left(\,M_1 \leftrightarrow M_2 \,\right),
 \label{Si3b}
\end{eqnarray}
where
$\Omega = (B-A)/(2k) $ in the above
expressions\footnote{Notice that if $(Q\,\theta/\beta)^2-n^2 <0$, we
  replace $\sin(...)$ by $\sinh(...)$ and
  $\sqrt{(Q\,\theta/\beta)^2-n^2}$ by $\sqrt{n^2-(Q\,\theta/\beta)^2}$.}.
We note that
$|\sigma_\textrm{r}^{\textrm{NP}}| \leq
|\sigma_\textrm{r}^{\textrm{P}}|$ for $\textrm{r} =0,a,b$.

From the above expressions, it is obvious that $\sigma^{\textrm{NP}}_0 \ra
\sigma_0^{\textrm{P}}$ when $k Q \, \theta \,\ll 1$.  In fact, to the
leading order, $Q \sim M_0$ where $M_0$ is the renormalized mass of
$\Phi$ (to be determined in the next subsection).  Thus, $ k Q \, \theta \ll
1 $ corresponds to the case where $ \theta \ll (1/M_0) \,(1/k)$. Notice
that $\sqrt{\theta}\equiv \lambda_{\textrm{NC}}$ represents the
characteristic length scale smaller than which the effect of 
noncommutative geometry
becomes significant. Moreover, the Compton and de-Broglie wavelengths
associated with $\Phi$ can be identified as $\la_{\textrm{C}} \sim
1/M_0$ and $\la_{\textrm{dB}}\sim 1/k$ respectively. As a result, we
conclude that $\sigma_0^{\textrm{NP}} \ra \sigma^{\textrm{P}}_0$ when $
\la_{\textrm{NC}} \ll \sqrt{\la_{\textrm{C}}\,
  \la_{\textrm{dB}}}$. This is always true when $\la_{\textrm{NC}}$ is
much smaller than the smaller of $\la_{\textrm{C}}$ and
$\la_{\textrm{dB}}$. In the relativistic case, $\la_{\textrm{dB}} \ll
\la_{\textrm{C}}$; while in the non-relativistic case,
$\la_{\textrm{C}} \ll \la_{\textrm{dB}}$. In any case, if
$\la_{\textrm{NC}}$ is really much smaller than the smaller of
$\la_{\textrm{C}}$ and $\la_{\textrm{dB}}$, then the resolution due to
either $\la_{\textrm{C}}$ or $\la_{\textrm{dB}}$ is not high enough to
see the effect of 
noncommutativity. The system behaves as if it were commutative,
and hence $\sigma_0^{\textrm{NP}} \ra \sigma^{\textrm{P}}_0$. On the
other hand, $\sigma_0^{\textrm{NP}} \ra 0$ when $k Q \,\theta \gg 1$, which
corresponds to the case $\la_{\textrm{NC}} \gg
\sqrt{\la_{\textrm{C}}\, \la_{\textrm{dB}}}$. This is always true when
$\la_{\textrm{NC}}$ is larger than the larger of $\la_{\textrm{C}}$
and $\la_{\textrm{dB}}$. In this case, the system is completely
noncommutative.

For $\sigma^{\textrm{NP}}_a$ and $\sigma^{\textrm{NP}}_b$, the significance
of 
noncommutativite effect depends on the ratio 
$Q\,\theta/\beta \sim M_0 T \, \theta$. When
$Q\,\theta/\beta \ll 1$, 
noncommutativity  is negligible.  This corresponds to
the case when $ \la_{\textrm{NC}} \ll  \la_{\textrm{T}}$ where $\la_{
  \textrm{T} } \sim 1/\sqrt{M_0 T}$ can be identified as the thermal
de-Broglie wavelength associated with $\Phi$ when it is propagating in
the thermal bath with temperature $T$. In the thermal medium, it is
the thermal de-Broglie wavelength that plays the role of the
characteristic resolution acquired by $\Phi$. When $\la_{\textrm{NC}}
\ll  \la_{\textrm{T}}$, the system behaves as if it were commutative,
and hence $\sigma^{\textrm{NP}}_{a,b} \ra \sigma^{\textrm{P}}_{a,b}$.  On
the other hand, the 
effect of noncommutative geometry 
is significant when $Q \,\theta/\beta \gg 1$,
which corresponds to $\la_{\textrm{NC}} \gg  \la_{\textrm{T}}$. In
this case, both $\sigma_a^{\textrm{NP}}$ and $\sigma_b^{\textrm{NP}}$
are suppressed.

In general,  as $\theta$ increases, both $\sigma_a^{\textrm{NP}}$  and
$\sigma_b^{\textrm{NP}}$ decrease. This can be attributed to the
reduction of degrees of freedom due to spacetime noncommutativity.
In particular, as $\theta\ra \infty$, $\sigma_{a,b}^{\textrm{NP}}\ra
0$. In this case, the reduction of degrees of freedom due to spacetime
noncommutativity is maximal. This can also be understood from the
extremely rapid oscillations of the phase factor in the integrands of
$\sigma_a^{\textrm{NP}}$ and $\sigma_b^{\textrm{NP}}$.

Furthermore, similar to the commutative case,  $\beta k \ll 1$
corresponds to the high temperature limit while $\beta k \gg 1$
corresponds to the low temperature limit. This is true regardless of
the relative significance of the 
noncommutative geometry.  Both
$\sigma_a^{\textrm{NP}}$ and $\sigma_b^{\textrm{NP}}$ increase with
decreasing $\beta k$. For instance, $\sigma^{\textrm{NP}}_{a,b} \ra 0$
when $\beta k \ra \infty$. On the other hand, both
$\sigma_a^{\textrm{NP}}$ and $\sigma^{\textrm{NP}}_b$ approach to a
finite value when $\beta k \ra 0$.

\subsection{Dispersion relation}

The real part of the self-energy is given by
\bea
\textrm{Re}\,\Sigma_\rR(\nu_{n}, \vec{k})= -g^{2}\int
\frac{d^{3}\vec{p}}{(2\pi)^{3}}\frac{1}{\beta}\sum_{\omega_{m}}
\frac{1}{\vec{p}^{2}+M_{1}^{2}+\om_m^2}
\frac{1}{(\vec{p}+\vec{k})^{2}+M_{2}^{2}+(\,\om_m+\nu_n\,)^2}
\big(\frac12 +\frac12 \,e^{ip\times k}\big). \nonumber
\eea
Again, the factors $\frac12$ and $\frac12 \,e^{ip\times k}$
represent the planar and non-planar contributions respectively.  To
facilitate the calculation, we introduce the Schwinger parameters
\bea
\frac{1}{\vec{p}^{2}+M_{1}^{2}+\om_m^2}&=&\int_{0}^{\infty}\,
d\al_1 \, e^{-\al_1\, (\,\vec{p}^{2}+M_{1}^{2}+\om_m^2\,)} \, ,
\\ \frac{1}{(\vec{p}+\vec{k})^{2}+M_{2}^{2}+(\,\om_m+\nu_n\,)^2}
&=&\int_{0}^{\infty}\, d\al_2 \, e^{-\al_2\,
  [\,(\vec{p}+\vec{k})^{2}+M_{2}^{2}+(\,\om_m+\nu_n\,)^2\,]}\,.
\eea

By completing squares, the $p$ integrals now becomes Gaussian and
can be readily evaluated to give
\bea \label{RR1}
\textrm{Re}\,\Sigma^{\textrm{P,\,NP}}_\rR(\nu_{n},
\vec{k}) &=&  -\frac{g^{2}}{64
  \pi^2}\,\int_{0}^{\infty}\,\frac{d\,\al}{\al}\,\,
e^{-\al\big[\,\frac{1}{4}\,k_E^2+\frac12 \,(\,M_1^2+M_2^2\,)\,\big]
  -\frac{L^2}{\al}} \nonumber \\
&& \qquad \cdot
\int_{-1}^{1}\,d\,x\,\,e^{-\frac{1}{4}\,\al \,k_E^2\,x^2
  -\frac12\,\al\,(\,M_1^2-M_2^2\,)x}\,\,\vartheta\left(\,\frac{n}{2}(1-x),\,
\frac{i\,\beta}{4\pi\al}\,\right), \; \;
\eea
where
$\al =\al_1+\al_2$, \;
$x = (\al_1-\al_2)/\al$, \;
$k_E^2 = k^2+\nu_n^2$, \,
and
\be
\vartheta(z,\tau) =\sum_{m=-\infty}^{\infty}\,
e^{2\pi i m z+ i  \pi m^2 \tau}
\ee
is the Jacobi theta function
\cite{WiseExtraNC}. Similar to \cite{SeibergPerturbative}, we have
multiplied the above integrands by $\exp{(-1/\al \,\Lambda^2)}$ in
order to regulate the small $\al$ divergence such that
\bea
L^2 = \left\{%
\begin{array}{ll}
    \dfrac{1}{\Lambda^2}, & \hbox{\,\,\textrm{for} \,
      $\textrm{Re}\,\Sigma^{\textrm{P}}_{\rm R}$;}
    \\
\dfrac{1}{\Lambda^2}+\dfrac{\tilde{k}^2}{4}, &
    \hbox{\,\,\textrm{for} \, $\textrm{Re}\,\Sigma^{\textrm{NP}}_{\rm R}$}
\\
\end{array}%
\right.   \, ,
\eea
where
$\tilde{k}= k\,\theta$.

The leading contribution of the integral \eq{RR1} comes from the
region $\alpha \sim 0$.  After performing the above integrations and
upon the analytic continuation $ \nu_n \ra -i \om -\epsilon$, we
obtain the retarded self-energy
$\textrm{Re}\,\Sigma_{\textrm{R}}(\om, k)
\equiv
I_0^{\textrm{P}}+I_0^{\textrm{NP}}+I^{\textrm{P}}+I^{\textrm{\,NP}}$
with
\bea
I_0^{\textrm{P,\,NP}} &=& -\frac{g^{2}}{32 \pi^2} \,2 \,
K_0\left(\,2 \,c^2\,L^2\,\right) \\
&\approx& \frac{g^{2}}{8 \pi^2} \,\ln\left(\,c\,L \,\right), \qquad
\qquad  \textrm{for} \,\, c\,L \ll 1, \\ I^{\textrm{P,\,NP}}
&=&\,-\frac{g^2\, Q^2}{8\pi^2
  (\beta\,\omega)^2}\sum_{m=1}^{\infty}\,\frac{1}{m^2}\,\left[\,
  \left(\,1+\frac{M_1^2-M_2^2}{Q^2}\,\right) \,\frac{L'}{M_1}\;
  K_1\left(\,2\,M_1\,L' \,\right)\nonumber \right. \\ &&
  \left. ~~~~~~~~~~~~~~~~~~~~~~~~~~~~~~~~ +
  \left(\,1-\frac{M_1^2-M_2^2}{Q^2}\,\right) \,\frac{L'}{M_2}\;
  K_1\left(\,2\,M_2\,L'\,\right)\,\right],   \label{IPNP}
\eea
\no
where
$Q^2=\omega^2-k^2 = -k_E^2$.  $K_0$ and $K_1$ are the modified Bessel
functions of the second kind.
Here
\be
c^2 = \frac{\big(Q^2 -(M_1-M_2)^2\big) \big((M_1+M_2)^2- Q^2\big)  }{4
  Q^2}
\ee
and
\be
L'{}^2 = L^2+\frac{m^2\beta^2}{4}.
\ee

In fact,
$I_0^{\textrm{P,\,NP}}$ and $I^{\textrm{P,\,NP}}$ arise from the $m=0$
and $m\neq 0 $ terms in the Jacobi theta function
respectively.
In the zero temperature limit $T\rightarrow 0$,
the only non-vanishing term in the Jacobi theta function comes from
$m=0$ in which case $\vartheta(z,\tau) \rightarrow 1$. Notice that $c$
is always positive-define if $(M_1-M_2)^2<Q^2< (M_1+M_2)^2$. For $Q^2>
(M_1+M_2)^2$, the $\Phi$ particle can decay into $\chi_1$ and
$\chi_2$. For $Q^2< (M_1-M_2)^2$, $\chi_1$ (or $\chi_2$) can decay
into $\Phi$ and $\chi_2$ (or $\chi_1$). In both cases, $c$ becomes
imaginary and a non-zero imaginary part will appear in
$I_0^{\textrm{P,\,NP}}$.

The bare mass $M_{B}$ of $\Phi$ receives renormalization from both of
$I_0^{\textrm{P}}$ and $I_0^{\textrm{NP}}$, and so the
(zero-temperature) renormalized mass for $\Phi$ is defined as
\be \label{Mpert}
M_0^2=M^2_{B}
+I_0^{\textrm{P}}|_{Q^2=M_0^2}+I_0^{\textrm{NP}}|_{Q^2=M_0^2}.
\ee
To
the order $\mathcal{O}(g^2)$, the dispersion relation is then given by
\bea \label{Q-d}
\om^2=k^2+M_0^2+\mathcal{I}^{\textrm{P}}
+\mathcal{I}^{\textrm{NP}},
\eea
\no
where
\be
\mathcal{I}^{\textrm{P}} =I^{\textrm{P}}|_{Q^2=M_0^2} \quad \mbox{and}
\quad \mathcal{I}^{\textrm{NP}} =I^{\textrm{NP}}|_{Q^2=M_0^2}
\ee
represent the finite-temperature corrections to the dispersion
relation.

It is instructive to examine the behaviour of the finite temperature
quantities $I^{\textrm{P, NP}}$ in various limits.  In the low
temperature limit  with $T \ll M_1, \,M_2$, we  have $M_{1,2} L'
\gg1$. Using $K_1 (x) \ra \sqrt{\frac{\pi}{2x}}\,e^{-x}$ for $x\gg1$,
it is obvious that $I^{\textrm{P,\,NP}}$ are exponentially suppressed
as long as $T \ll M_1,\, M_2$, regardless of the significance of the
noncommutativity. 
This is consistent with the fact that $I^{\textrm{P,\,NP}}$
arise from the $m\neq 0$ terms in the Jacobi theta function and can
survive only at finite temperature.  Similarly,  $I^{\textrm{NP}}$ is
exponentially suppressed in the large 
$\th$ limit with $1/k\theta \ll
M_1, \,M_2$, regardless of the magnitude of the temperature $T$.
On the other hand, in the high temperature limit with
 $M_1, \, M_2 \ll T, \, 1/k\theta$,
we have $M_{1,2} L' \ll 1$.
Using $K_1 (x) \ra \frac{1}{x}$ for $x\ll1$, we obtain
\bea
I^{\textrm{P,\,NP}}
&\approx& \,-\frac{g^2\, Q^2}{96 \,( \beta \omega)^2} \,\left[\,
  \left(\,1+\frac{M_1^2-M_2^2}{Q^2}\,\right) \,\frac{1}{M_1^2} +
  \left(\,1-\frac{M_1^2-M_2^2}{Q^2}\,\right) \,\frac{1}{M_2^2}
  \,\right] \label{IPNP_highT} \, .
\eea
It is remarkable to notice
that the above expression for $I^{\textrm{P,\,NP}}$ are independent of
$m$ and the ultraviolet cut-off $\Lambda$.  Also, it is valid
irrespective to the relative magnitude between $T$ and
$1/k\theta$. Most interestingly, the non-planar contribution
$I^{\textrm{NP}}$ is completely independent of $\theta$. This
means that in the limit $M_1, \, M_2 \ll T, \, 1/k\theta$, the
dispersion relation receives vanishing finite-temperature corrections
from the spacetime noncommutativity.

Finally, let us comment on the 
IR/UV mixing effect of the
noncommutative field  theory at finite temperature.  Needless to say,
since $I_0^{\textrm{NP}}$ is a zero-temperature non-planar quantity,
it does suffer from the usual IR/UV mixing.
Since the IR singularities are a reflection of the fact that the field
theory is UV divergent, the key to resolve the IR singularities lies
at a proper UV finite completion of the noncommutative field theory
\cite{ACG}. With
a choice of UV completion, the IR singularities
will get smoothen out.  For example, a natural choice is to embed the
noncommutative field theory as a low energy field theory of open
string theory in background $B$-fields. It was shown explicitly in \cite{ACG}
that in doing so, the IR pathologies of noncommutative
field theory are resolved. In particular, in the deep IR, the theory
flows continuously to the commutative field theory and the normal
Wilsonian behaviour is restored. Therefore, with this understanding,
the zero-temperature IR/UV singularities are harmless.

It is clear that our noncommutative thermal bath does
not suffer from any further IR/UV mixing problem, in the sense that
all the finite-temperature non-planar quantities are healthy and are
absent from any infrared singularities if we take the limit
$k\rightarrow 0$ after taking the limit $\Lambda \rightarrow
\infty$. For instance, let us look at the quantity $L'$ in
$I^{\textrm{NP}}$ (which is of purely finite-temperature nature) as
displayed in (\ref{IPNP}). It is obvious that $L'{}^2\rightarrow
m^2\beta^2/4$ when $\Lambda \rightarrow \infty$ and $k\rightarrow
0$. Unless $T\rightarrow \infty$, $I^{\textrm{NP}}$ is manifestly
finite.
Physically, it is the finite temperature $T$ acquired by the thermal
bath that rescues the system from any further IR/UV mixing problem:
the finite temperature $T$ acts as an effective ultraviolet cut-off
for the system once we have taken the limit $\Lambda \rightarrow
\infty$.

\section{Effects of Noncommutativity on
Nonequilibrium Dynamics}

In previous section, we have computed the imaginary part of the
self-energy as well as the dispersion 
relation associated with the
$\Phi$ 
particles. 
Now, we can proceed to study the nonequilibrium
dynamics of $\Phi$ when it propagates in the noncommutative
spacetime. In particular, we will focus on how does spacetime
noncommutativity affect the possible decay and thermalization
processes of $\Phi$ in the thermal bath.
Without loss of generality, we will assume that $M_1 \geq  M_2$.

\subsection{Decay and thermalization rates}

Let us write the imaginary part of the self-energy as
$\text{Im}\Sigma_\textrm{R} = \sigma_{\textrm{D}}+\sigma_{\textrm{T}}$
where $\sigma_{\textrm{D}} =
\sigma^{\textrm{P}}_0+\sigma^{\textrm{P}}_a
+\sigma_0^{\textrm{NP}}+\sigma_a^{\textrm{NP}} $
and $\sigma_{\textrm{T}}=\sigma^{\textrm{P}}_b+\sigma_b^{\textrm{NP}}$.
These quantities are regulated by kinematical constraints and take
the forms:
\bea
\sigma_{\textrm{D}} &=& \Theta[\,Q^2-(M_{1}+M_{2})^{2}\,] \;
\Lambda_{\textrm{D}} (Q^2, k^2), \\
\sigma_{\textrm{T}} &=& \Theta[\,(M_{1}-M_{2})^{2}-Q^2\,] \;
\Lambda_{\textrm{T}} (Q^2, k^2) +\Theta(\,-Q^2\,) \;
\Lambda'_{\textrm{T}} (Q^2, k^2),
\eea
where $\Lambda_{\textrm{D}} (Q^2, k^2)$ can be read off from \eq{Si1a},
\eq{Si2a}, \eq{Si1b}, \eq{Si2b},
while $\Lambda_{\textrm{T}} (Q^2, k^2)$ and $\Lambda'_{\textrm{T}} (Q^2, k^2)$
can be read off from \eq{Si3a},
 \eq{Si3b} respectively.
To compute the decay and thermalization rates \cite{CMHo,lebellac},
we need to put the
$\Phi$ field on-shell, i.e. setting
\be \label{QII}
Q^2 =M_0^2 + \cI^\rP + \cI^{\rNP}.
\ee
Then, to
the order $\mathcal{O}(g^2)$, we obtain
\be
\Gamma_{\textrm{D}}
= \frac{\Lambda_{\textrm{D}}(M_0^2, k^2) }{2\,\om_0} \;
\Theta[\,Q^2-(M_{1}+M_{2})^{2}\,]
\ee
and
\be
\Gamma_{\textrm{T}}
=
\frac{\Lambda_{\textrm{T}}(M_0^2, k^2)}{2\,\om_0}  \;
\Theta[\,(M_{1}-M_{2})^{2}-Q^2\,],
\ee
where
$\om_0=\sqrt{k^2+M_0^2}$. Note that up to the order
$\mathcal{O}(g^2)$, it is sufficient to set
$Q^2= M_0^2$ in  $\Lambda_\rD$ and $\Lambda_\rT$;
however, one has to use the full expression \eq{QII}
inside the $\Theta$ functions.

Physically,
$\sigma^{\textrm{P}}_0+\sigma_0^{\textrm{NP}}$ represents the
zero-temperature planar $+$ non-planar contributions to the decay
rate, while $\sigma^{\textrm{P}}_a+\sigma_a^{\textrm{NP}}$ represents
the finite-temperature planar $+$ non-planar contributions to the
decay rate. Similarly, $\sigma^{\textrm{P}}_b+\sigma_b^{\textrm{NP}}$
represents the finite-temperature planar $+$ non-planar contributions
to the thermalization rate. Notice that
$\sigma^{\textrm{P}}_b+\sigma_b^{\textrm{NP}}$ is a purely
finite-temperature effect and there is no zero-temperature counterpart
of it, namely $\sigma^{\textrm{P}}_b+\sigma_b^{\textrm{NP}} \rightarrow 0 $
as $T\rightarrow 0$.

The rates $\Gamma_\rD(k)$ and  $\Gamma_\rT(k)$ depend on $k$,
the magnitude of the spatial momentum.
If $\Gamma_\textrm{D}(k) =0$ for a certain $k$, then
this mode of the $\Phi$ particle is stable in the
thermal medium. Conversely, if $\Gamma_\textrm{D}(k)\neq 0$
for a certain $k$,
then this mode
would be able to decay into $\chi_1+\chi_2$. Kinematically, the
feasibility of decay is regulated by the Heaviside function
\be \label{ar1}
\Theta[\,Q^2-(M_{1}+M_{2})^{2}\,].
\ee
On the other hand,
$\Gamma_{\textrm{T}} (k)\neq 0$ for a certain $k$ implies that this mode of
$\Phi$ acquires a relaxation
or thermalization time scale beyond which it approaches thermal
equilibrium with the bath constituted by $\chi_1$ and $\chi_2$. This
happens through the decay of $\chi_{1,2}$ into $\chi_{2,1}$ and
$\Phi$, and their recombination, namely $\chi_{1,2} \leftrightarrow
\chi_{2,1} +\Phi$ \cite{CMHo,lebellac}. As the
modes carrying momentum number $k$ propagate through the thermal bath,
they will be screened or dressed
by the excitations in the medium and
will propagate as quasi-particles.
In fact, $\Gamma_\textrm{T}(k)$ characterizes the
``decay rate" of the quasi-particles associated with $\Phi$ in the
medium, and this is precisely the relaxation or thermalization rate
\cite{CMHo,lebellac}.  Of course, if $\Gamma_\textrm{T}(k) = 0$, then it
takes an infinitely long time for
these modes of $\Phi$
to approach thermal
equilibrium with the bath, which simply means that
these modes
can never
thermalize with the bath. Kinematically, the feasibility of thermalization
is regulated by the Heaviside function
\be \label{ar2}
\Theta[\,(M_{1}-M_{2})^{2}-Q^2\,].
\ee


\subsection{Momentum dependent alternation of stability and
thermalizability}

What we would like to study is the following.
How does the conventional understanding of the
dynamics of $\Phi$ in the thermal bath change
in the presence of noncommutative geometry?

Since the decay and thermalization processes depend crucially on
  the relative size of $Q^2$  compared to $(M_1-M_2)^2$ and
  $(M_1+M_2)^2$, it is useful to consider the following three regions
  of $Q^2$:
\bea
\cR_1&:& \, Q^2<(M_1-M_2)^2, \quad
\mbox{where we have $\Gamma_{\textrm{T}} \neq 0$ and
  $\Gamma_{\textrm{D}}  =0$}, \nn\\
\cR_2&:&\, (M_1-M_2)^2<Q^2 <(M_1+M_2)^2, \quad \mbox{where we have
$\Gamma_{\textrm{T}} =  \Gamma_{\textrm{D}} =0$},\nn\\
\cR_3&:& \, (M_1+M_2)^2< Q^2, \quad
\mbox{where we have $\Gamma_{\textrm{T}} = 0$ and
  $\Gamma_{\textrm{D}} \neq 0$}.
\eea
If we define
$Q_\textrm{C}^2$ as the commutative counterpart of $Q^2$, then it
would be given by $Q_\textrm{C}^2=M_0^2+
2\,\mathcal{I}^{\textrm{P}}$.
When  one turns  on the noncommutativity parameter,  $Q_\rC^2$ changes
to $Q^2$  and
  could cross over from one region to another, corresponding to
  turning on or turning off the decay and/or the  thermalization
  processes as one takes into account of the effects of
  noncommutativity. What can actually happen depends on the ``jump''
$Q^2-Q^2_{\textrm{C}}=\mathcal{I}^{\textrm{NP}}-\mathcal{I}^{\textrm{P}}$.
Since $|\mathcal{I}^{\textrm{NP}}|\leq |\mathcal{I}^{\textrm{P}}|$, we
will always have $Q^2-Q^2_{\textrm{C}} <0$ \,if\,
$\mathcal{I}^{\textrm{P}}>0$\,
and\,
$Q^2-Q^2_{\textrm{C}} >0$ \,if \,$\mathcal{I}^{\textrm{P}}<0$, regardless
of the signs of $\mathcal{I}^{\textrm{NP}}$.
Therefore the  direction of the ``jump'' depends solely on the
sign of $\cI^\rP$. By setting
$\Lambda \rightarrow \infty$ and $Q^2=M_0^2$ in (\ref{IPNP}), it is
easy to  obtain that
\bea \label{c1}
&&\mathcal{I}^{\textrm{P}}>0 \quad
\mbox{when} \quad M_0^2 <\frac{(M_1^2-M_2^2)^2}{M_1^2+M_2^2},
\\
\label{c2}
&&\mathcal{I}^{\textrm{P}}<0 \quad \mbox{when} \quad M_1^2-M_2^2 <
M_0^2
\eea
for any temperature $T$ and momentum $k$.
And for $M_0^2$ in the intermediate region,
one obtains
\begin{empheq}[right=\empheqbigrbrace
\quad \mbox{when} \quad \dfrac{(M_1^2-M_2^2)^2}{M_1^2+M_2^2}
\leq M_0^2 \leq  M_1^2-M_2^2. ]{align}
\cI^\rP &>0 \quad \mbox{for $T< T_0$}, \label{c3}\\
\cI^\rP &<0 \quad \mbox{for $T > T_0$}, \nn
\end{empheq}
Here  $T_0 $ is a temperature that is
determined by
solving $\cI^\rP(T_0)=0$.
Note that $T_0$ depends not just on the mass $M_0$ of
the propagating particles $\Phi$ and the
properties of the thermal bath (i.e. $M_1, M_2$),
but also on the momentum $k$ of the
mode of $\Phi$ in
consideration.
Therefore, as we remarked above,
the impact of spacetime noncommutativity on the decay and
thermalization processes is going to be momentum dependent.
We remark that due to the rather complicated form for $\cI^\rP$, an
analytic expression for $T_0$ is not available. Fortunately, this is
actually not needed in our analysis below which is aimed at
explaining the general physical features.
In a concrete phenomenological study,  a more detailed knowledge
of $T_0$ maybe needed and this can always be obtained numerically.

As we mentioned above, the decay and
thermalization processes may be turned on or off when one takes into account of
the effects of noncommutativity. This happens whenever $Q^2$ crosses
over different regions as noncommutativity is turned on. As a result, we obtain
the following six interesting  cases which may happen
in principle.

\vskip 0.3cm
\no \underline{\bf 1. $Q_{\textrm{C}}^2$ in region $\cR_1$:
~ $Q_{\textrm{C}}^2<(M_1-M_2)^2$}
\vskip 0.2cm

\no In the commutative
picture, we have $\Gamma_\textrm{D}=0$ and
$\Gamma_\textrm{T}\neq0$ which imply that $\Phi$ is stable and cannot
decay into $\chi_1+\chi_2$, but it can thermalize with the bath. Once
we switch to the noncommutative bath,
we may have modes such that

\begin{itemize}

\item{Case A: ~ $(M_1-M_2)^2 < Q^2 < (M_1+M_2)^2 ~\Longrightarrow
  ~\Gamma_\textrm{D}=\Gamma_\textrm{T} = 0 $.  This corresponds to the
  situation with $\mathcal{I}^{\textrm{P}}<0$. These modes of $\Phi$
  are still stable against decay into $\chi_1+\chi_2$, but they cannot
  thermalize with the noncommutative bath anymore.
In this case , we see  that 
noncommutative geometry
suppresses thermalization for these modes.}

\item{Case B: ~ $(M_1+M_2)^2< Q^2 ~\Longrightarrow ~ \Gamma_\textrm{D}
  \neq 0$ and $\Gamma_\textrm{T} = 0$. This also corresponds to the
  situation with $\mathcal{I}^{\textrm{P}}<0$.  These modes of $\Phi$
  which are originally stable in the commutative bath, can now decay
  into $\chi_1+\chi_2$ in the noncommutative bath, but they cannot
  thermalize with it anymore. In this case, 
noncommutative geometry 
induces  decay but suppresses thermalization for these modes.  }

\end{itemize}

\no \underline{\bf
2. $Q_{\textrm{C}}^2$ in region $\cR_2$:
~ $(M_1-M_2)^2<Q_{\textrm{C}}^2<(M_1+M_2)^2$}
\vskip 0.2cm

In the commutative picture, we have $\Gamma_\textrm{D}=\Gamma_\textrm{T}
= 0$ which implies that $\Phi$ can neither decay into $\chi_1+\chi_2$
nor thermalize with the bath. Once we switch to the noncommutative
bath, we may have modes  such that

\begin{itemize}

\item{Case C: ~ $0< Q^2 < (M_1-M_2)^2 ~\Longrightarrow ~
  \Gamma_\textrm{D} = 0$ and $\Gamma_\textrm{T} \neq 0 $. This
  corresponds to the situation with $\mathcal{I}^{\textrm{P}}>0$.
  These modes of $\Phi$ still cannot decay into $\chi_1+\chi_2$, but
  they can now thermalize with the noncommutative bath. In this case, 
noncommutative geometry  induces thermalization for these modes.}

\item{Case D: ~ $(M_1+M_2)^2< Q^2 ~\Longrightarrow ~ \Gamma_\textrm{D}
  \neq 0$ and $\Gamma_\textrm{T} = 0 $. This corresponds to the
  situation with $\mathcal{I}^{\textrm{P}}<0$. These modes of $\Phi$
  can now decay into $\chi_1+\chi_2$ in the noncommutative bath, but
  still cannot thermalize with it. In this case, 
noncommutative geometry  induces decay for  these modes.}

\end{itemize}

\no \underline{\bf
3. $Q_{\textrm{C}}^2$ in region $\cR_3$:
~ $(M_1+M_2)^2<Q^2_{\rm C}$}
\vskip 0.2cm

In the commutative picture, we have $\Gamma_\textrm{D} \neq 0$ and
$\Gamma_\textrm{T} = 0$ which imply $\Phi$ is unstable and can decay
into $\chi_1+\chi_2$, but it cannot thermalize with the bath. Once we
switch to the noncommutative bath,  we may have modes such that

\begin{itemize}

\item{Case E: ~ $0< Q^2 < (M_1-M_2)^2 ~\Longrightarrow ~
  \Gamma_\textrm{D} = 0$ and $\Gamma_\textrm{T} \neq 0$. This
  corresponds to the situation with $\mathcal{I}^{\textrm{P}}>0$.
  These modes of $\Phi$  cannot decay into $\chi_1+\chi_2$ in the
  noncommutative bath anymore, but they can now thermalize with it. In
  this case,
noncommutative geometry suppresses decay but induces  thermalization for these
  modes.}

\item{Case F: ~ $(M_1-M_2)^2 < Q^2 < (M_1+M_2)^2 ~\Longrightarrow ~
  \Gamma_\textrm{D}=\Gamma_\textrm{T} = 0$. This also corresponds to
  the situation with $\mathcal{I}^{\textrm{P}}>0$.  These modes of
  $\Phi$ cannot decay into $\chi_1+\chi_2$ in the noncommutative bath
  anymore, and they still cannot thermalize with it.  In this case, 
noncommutative geometry   suppresses decay for these modes.}

\end{itemize}

For a given thermal bath and a propagating particle, the set of
masses $M_0$, $M_1$, $M_2$ and the coupling $g$ are fixed.
To decide which of the above six cases can occur,
one need to examine carefully how  $Q^2$ changes
when noncommutativity is turned on.
As an illustration,
we present
the detailed analysis of cases A and B in the appendix A.
The result is that cases A and B can occur only if
the masses of the thermal bath satisfy
\be \label{cc7-0}
\frac{M_1}{M_2} > 4+ \sqrt{15}.
\ee
and if $M_0$
falls in the range of
\be \label{cc6-0}
\frac{(M_1^2-M_2^2)^2}{M_1^2+M_2^2}< M_0^2 < \frac{5}{4}(M_1-M_2)^2.
\ee
When these conditions are satisfied,
then depending on the momentum $k$ of the mode of concern,
either case A or case B or
both cases can occur. See \eq{d1}-\eq{d7}.

Similar analysis can be performed for the other four cases. The
conclusion is similar. For cases E and F to occur, the necessary
conditions are that the masses of thermal bath has to satisfy
\be
M_1/M_2 > 11
\ee
and $M_0$ has be in the range
\be \label{M0-M12}
\frac{5}{6} (M_1+M_2)^2 < M_0^2 <M_1^2-M_2^2.
\ee
For cases C and D,  no condition is needed on
$M_1/M_2$. But   \eq{M0-M12} is needed
for the case C and \eq{cc6} is needed for the case D to occur. Similar
conditions as \eq{d1}-\eq{d7} can be written down for these four cases.
We  will skip them here.

All in all, we conclude that all of the six cases A-F listed above can
occur in general. Depending on the masses, temperature of the thermal
bath and the magnitude $k$ of the spatial momentum of the mode of concern,
the decay and thermalization rates of
$\Phi$ can be either induced or suppressed when
one takes into account of noncommutativity properly.
In general,
there is a window of momentum modes
whose stability and/or thermalizability properties are altered.
The window
depends on the temperature of the thermal bath. Typically, the affected modes
are in the high (respectively low) $k$ regime when $T$ is low
(respectively high).


\section{Conclusions}

In this article, we studied the nonequilibrium dynamics of a scalar
$\Phi$ propagating through a noncommutative thermal bath.  We
showed that
noncommutative geometry has a distinct impact on
the nonequilibrium dynamics of particles propagating in a thermal bath
by providing a momentum dependent
enhancement or suppression of
their decay or thermalization processes. This is a combined effect of
finite temperature and noncommutative geometry.
Also, we pointed out that the finite temperature $T$
of the thermal bath can play the role as an effective ultraviolet
cut-off which rescues all the finite-temperature non-planar quantities
from any further IR/UV mixing problem.

Although our thermal bath is represented in a specific
way by the two scalar particles $\chi_1$ and $\chi_2$,
our analysis is general and our conclusions should
apply {generally for any bosons in contact with a noncommutative thermal bath.

A particularly interesting arena of application of our results
is the early universe at which spacetime noncommutativity may be present.
Our  results call for a re-examination of some of the
important processes in the early universe such as
reheating after inflation, baryogenesis and the freeze-out
of superheavy dark matter candidates, which are generally believed to occur
at very high energy scales where spacetime noncommutativity
could be significant
enough. It will be very interesting to re-examine these processes in details and
see how the
presence of noncommutative geometry could affect the conventional picture.
This could also provide
another channel for probing the presence of
noncommutative geometry in the early universe, in addition to the
possible cosmological imprint left by inflation, see for example
\cite{inf1,inf2,inf3}.

\section*{Acknowledgements}
C. S. Chu would like to thank the Berkeley Center for Theoretical Physics
for hospitality during his visit.
C. M. Ho would like to thank Ori Ganor for useful discussions.
The work of C. S. Chu has been supported by EPSRC and STFC.

\appendix

\section{Analysis of cases A and B}

Modes for which the cases A and B  can occur
have to satisfy the following
conditions:

\bea
&M_0^2 + 2 \cI^\rP &< (M_1-M_2)^2, \label{cc1}\\
&|\cI^\rP| &< 0.1 M_0^2, \label{cc2}\\
(M_1 -M_2)^2 <& M_0^2 + \cI^\rP + \cI^\rNP &< (M_1 +M_2)^2, \label{cc3a}\\
(M_1 +M_2)^2 <& M_0^2 + \cI^\rP + \cI^\rNP&. \label{cc3b}
\eea
Here the first condition specifies that one is initially in the region
$\cR_1$. The second condition \eq{cc2}
is a condition on the  size of the thermal correction to $M_0^2$.
We have imposed a conservative  $10 \% $ correction so as
to guarantee that we are in the validity regime of the
perturbation theory.  In principle,
one could have taken a different value for the RHS of \eq{cc2},
e.g. $0.2M_0^2$.
This will not modify the analysis below and the physical effects we
are going to point out are generic. The condition \eq{cc3a} is
for case A and the condition \eq{cc3b} is for case B.

To analyze these conditions, we first note that
cases A and B can occur only if $\cI^\rP<0$. This means that it is
necessary to require
$ M_0^2 > (M_1^2-M_2^2)^2/(M_1^2+M_2^2)$. On the other hand,
the
conditions \eq{cc1} and \eq{cc2} imply
that
\be \label{cc5}
\frac{1}{2}[M_0^2 -(M_1-M_2)^2] < -\cI^\rP < 0.1 M_0^2.
\ee
This is a non-empty condition for $\cI^\rP$ only if
$M_0^2 < \frac{5}{4}(M_1-M_2)^2$. This condition, together with the
previous condition on $M_0^2$, gives
\be \label{cc6}
\frac{(M_1^2-M_2^2)^2}{M_1^2+M_2^2}< M_0^2 < \frac{5}{4}(M_1-M_2)^2.
\ee
This specifies a non-empty range for $M_0^2$
only if $ (M_1^2-M_2^2)^2/(M_1^2+M_2^2)<
\frac{5}{4}(M_1-M_2)^2$, i.e.
\be \label{cc7}
\frac{M_1}{M_2} > 4+ \sqrt{15}.
\ee
The condition \eq{cc7} on the ratio $M_1/M_2$ and the condition \eq{cc6} on
$M_0$ are common to cases A and B and are the necessary conditions
for the cases A and B to occur.
Now assume \eq{cc7} and \eq{cc6} are  satisfied,
we are guaranteed that \eq{cc5} specifies a non-empty range for
$- \cI^\rP$. The question is whether and for what configuration of modes
will \eq{cc5} be satisfied. To address this, we need
the form of $\cI^\rP$ as a
function of $k$ and $T$,
which we collect here (also for $\cI^\rNP$):
\bea
\cI^\rP = -\frac{g^2 M_0^2}{16\pi^2 \o^2} \,T^2
\,\sum_{m=1}^\infty
\bigg[
\frac{c_+}{m^2 \,M_1^2}\, h(m \,x_1) + \frac{c_-}{m^2 \,M_2^2}\,
h(m \,x_2) \bigg],\\
\cI^\rNP = -\frac{g^2 M_0^2}{16\pi^2 \o^2} \,T^2
\,\sum_{m=1}^\infty
\bigg[
\frac{c_+}{m^2\, M_1^2} \,h(m \,y_1) + \frac{c_-}{m^2 \,M_2^2}\,
h(m\, y_2) \bigg].
\eea
Here $c_\pm = 1 \pm (M_1^2-M_2^2)/M_0^2$,
\be
x_i = M_i /T , \quad y_i = \sqrt{1/T^2+ k^2 \th^2}\, M_i, \quad i=1,2
\ee
and
\be
h(x) : = x \,K_1(x).
\ee
To analyze which set of modes satisfy the condition \eq{cc5}, we note
that
$- \cI^\rP =0$ at $T=0$ and $ - \cI^\rP \sim d \times T^2$
for large $T \gg M_1, M_2$, where
$d \approx g^2 M_0^2/(96 \o^2)
[c_+/M_1^2 + c_-/M_2^2 ] >0$. Therefore, for
any given $k$, there is a temperature range
\be \label{cc-T1}
T_1(k) < T< T_2(k)
\ee
such that \eq{cc5} is satisfied. Here the temperatures $T_{1,2}(k)$ are
determined by solving
\be
\cI^\rP(k,T_1(k)) = \frac{1}{2}[M_0^2
-(M_1-M_2)^2] \quad \mbox{and} \quad
\cI^\rP(k,T_2(k)) =  0.1 M_0^2.
\ee

\psfrag{0}{$0$}
\psfrag{Ik}{$-\cI^\rP(k)$, $k$ fixed}
\psfrag{I0}{$-\cI^\rP \sim T^2$}
\psfrag{B}{$0.1 M_0^2$}
\psfrag{A}{$\frac{1}{2}(M_0^2-(M_1-M_2)^2)$}
\psfrag{T}{$T$}
\psfrag{T1}{$T_1(k)$}
\psfrag{T2}{$T_2(k)$}
\begin{figure}[bpt]
\label{fig2}
\begin{center}
\includegraphics[scale=.7]{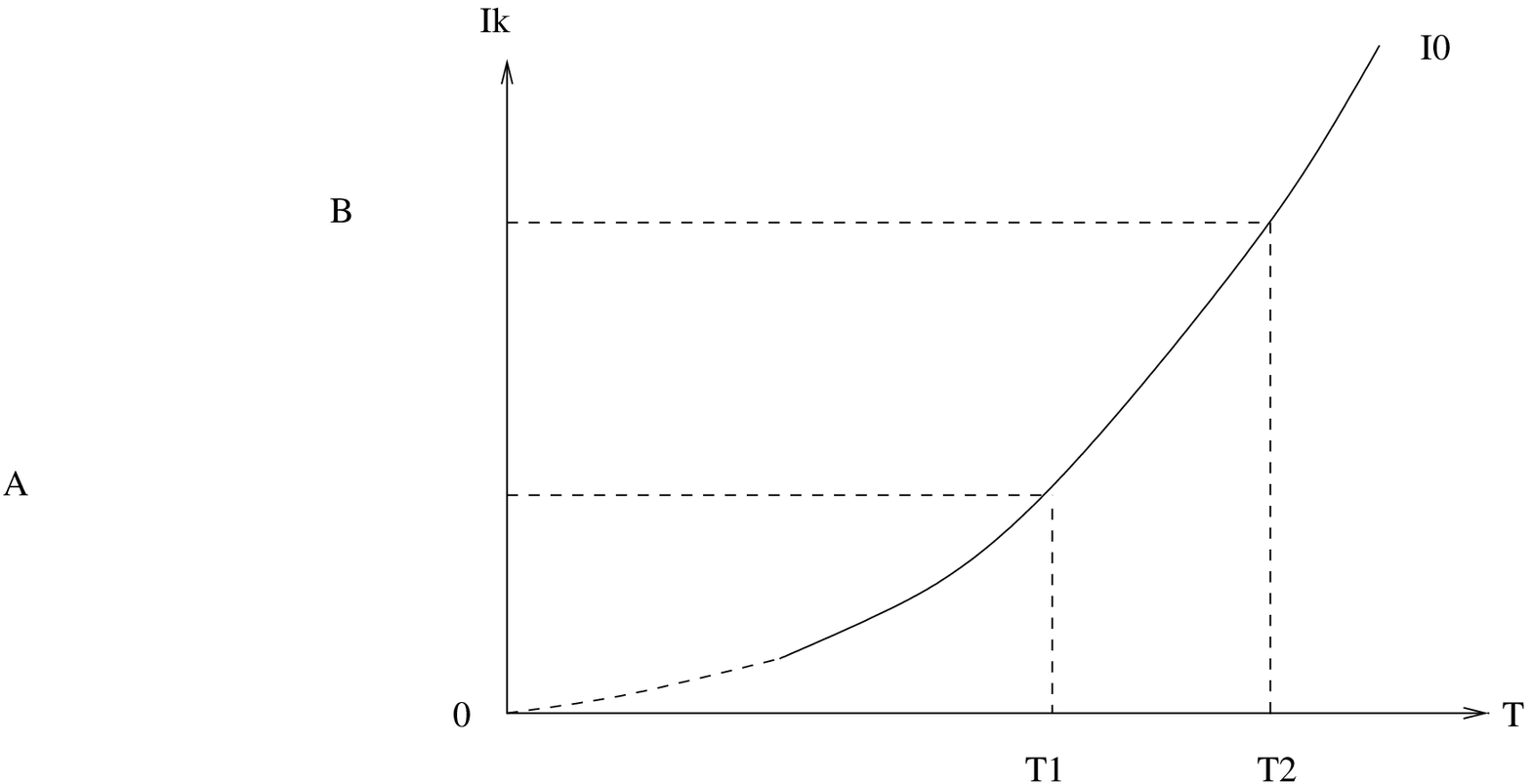}
\\{Fig 2: Solving the condition \eq{cc5}. }
\end{center}
\end{figure}

So far the obtained conditions \eq{cc7}, \eq{cc6} and \eq{cc-T1} are common
to both case A and case B.
Let us now proceed to analyze the conditions \eq{cc3a} and \eq{cc3b}.
Define the temperatures $\Tt_{1,2}$ by
\be
-(\cI^\rP + \cI^\rNP)(k,\Tt_1(k)) = M_0^2 -(M_1+M_2)^2
\quad \mbox{and} \quad
-(\cI^\rP + \cI^\rNP)(k,\Tt_2(k)) = M_0^2 -(M_1-M_2)^2,
\ee
then for a mode with momentum $k$ in a thermal bath with  temperature
$T$,
\bea
\mbox{case A occurs if} && \Tt_1(k) < T < \Tt_2(k), \label{cc-T2} \\
\mbox{case B occurs if} && \quad \qquad T< \Tt_1(k) . \label{cc-T3}
\eea
Now since $-(\cI^\rP +\cI^\rNP) < -2
\cI^\rP$, it follows that $\Tt_2(k) > T_1(k)$ for all
$k$. This is important since it implies that the condition \eq{cc-T1}
has always a non-trivial intersection with the conditions \eq{cc-T2}
or \eq{cc-T3}. For each mode with a fixed momentum $k$,
depending on the relative sizes of $T_1(k), T_2(k),
\Tt_1(k), \Tt_2(k)$, either case A or case B or both cases can occur.
Specifically we obtain the following result:

\noindent
When $\Tt_2(k) < T_2(k)$:
\bea
\mbox{if $\Tt_1(k) < T_1(k)$, then} \quad && \mbox{case A occurs for
  $T_1(k) <T<\Tt_2(k)$}; \label{d1}\\
\mbox{if $\Tt_1(k) > T_1(k)$, then} \quad && \mbox{case A occurs for
  $\Tt_1(k) <T<\Tt_2(k)$}; \\
&& \mbox{case B occurs for  $T_1(k) <T<\Tt_1(k)$}.
\eea
\noindent
When $\Tt_2(k) > T_2(k)$:
\bea
\mbox{if $\Tt_1(k) > T_2(k)$, then} \quad && \mbox{case B occurs for
  $T_1(k) <T< T_2(k)$}; \\
\mbox{if $T_1(k)< \Tt_1(k) < T_2(k)$, then} \quad && \mbox{case A occurs for
  $\Tt_1(k) <T<T_2(k)$}; \\
&& \mbox{case B occurs for  $T_1(k) <T<\Tt_1(k)$};\\
\mbox{if $\Tt_1(k) < T_1(k)$, then} \quad && \mbox{case A occurs for
  $T_1(k) <T<T_2(k)$}. \label{d7}
\eea

This means that given the temperature $T$ of the thermal bath, modes with
different (spatial) momentum $k$ can behave quite differently in regard to the
thermalization or stability properties. In particular only a certain
window of momentum modes as specified by the inequalities in \eq{d1}-
\eq{d7} above will be affected for a given $T$.
Since the characteristic
temperatures  $T_{1,2}(k), \Tt_{1,2}(k)$ are monotonic decreasing
function of $k$, it means for low temperature $T$, the affected window
is in the higher $k$ regime; while for high temperature $T$, the affected window
is in the lower $k$ regime.
This momentum dependent
alternation of the
stability or thermalizability is novel and is a combined effect
of finite temperature and noncommutative geometry.

\end{document}